%% file: seyfert.tex
\documentstyle[psfig]{l-aa}

\def\lapp{\mathbin{\raise2pt \hbox{$<$} \hskip-9pt \lower4pt
\hbox{$\sim$}}}
\def\gapp{\mathbin{\raise2pt \hbox{$>$} \hskip-9pt \lower4pt
\hbox{$\sim$}}}

\begin{document}

\thesaurus{02.08.1; 09.10.1; 11.19.1}

\title{Understanding NLR in Seyfert Galaxies: numerical simulation of 
jet-cloud interaction}
\subtitle{} 

\author{ P.\ Rossi\inst{1}\and A.\ Capetti \inst{1} \and G.\ Bodo\inst{1} 
\and  S.\ Massaglia \inst{2} \and A.\ Ferrari\inst{1,2}} 

\institute{Osservatorio Astronomico di Torino, I-10025 Pino Torinese, Italy\\
email: rossi@to.astro.it, capetti@to.astro.it, bodo@to.astro.it
\and Dipartimento di Fisica Generale dell'Universit\`a, Via Pietro
Giuria 1, I-10125 Torino, Italy\\
email: massaglia@ph.unito.it, ferrari@ph.unito.it} 

\offprints{P. Rossi}
%
\date{Received / Accepted}
%
\maketitle
\begin{abstract}
Recent HST observations suggest that the NLR in Seyfert Galaxies can be the 
result 
of interaction between jet and external inhomogeneous medium; following this 
suggestion we 
perform numerical simulations considering the impact of a radiative
jet on a dense cloud. We approach the problem adopting  a 
hydrodynamical code, that consents us to study in detail  
the jet hydrodynamics, while we choose a more simplified
treatment of 
radiative  processes, in order to give a qualitatively good interpretation of 
the emission processes.  Our three main purposes are: 
i) to reproduce in our simulations the physical conditions observed in the NLR
of Seyfert Galaxies, ii) to obtain 
physical constraints of the jet parameters and iii) to study the jet 
capacity to photoionize the surrounding medium.

We find that the jet-cloud interaction leads to clumps of matter with
density, temperature and velocity that agree with observations.
Conversely, the 
photoionizing flux radiated by the jet-induced shocks  
does not appear to be sufficient to account for the NLR line luminosity
but it may produce local and transient effects on the NLR ionization balance.

Finally, the observational requirements can be matched only
if jets in Seyfert galaxies are relatively heavy,
$\rho_{\rm jet}\, \gapp 1 $ cm $^{-3}$, and with velocities smaller than  
$\lapp \, 50,000\,$km s$^{-1}$, very different from 
their counterparts in radio-galaxies.

\end{abstract}
\thesaurus{}
\section{Introduction}

Extensive HST emission-line imaging  
of Seyfert galaxies has for the first time
resolved details of the structure of their Narrow Line Regions (NLR). 
In several cases cone-like morphologies have been revealed, 
similar in shape to - but of much smaller linear extent than - 
the Extended Narrow Line Regions (ENLR) seen
in the lower resolution ground based images 
(Wilson \& Tsvetanov, 1994 and references therein). 
In the standard model of the NLR, the UV 
emission of the nucleus is responsible of photoionizing the
Interstellar Medium (ISM) of the host galaxy. 
These conical distributions of the ionized gas 
have been interpreted as a confirmation
of the anisotropy of the nuclear radiation field which, in the framework 
of the unified scheme for Seyfert galaxies (e.g. Antonucci 1993) is caused by
the shadowing of an obscuring circumnuclear torus. 
However, in galaxies with linear radio structures,
the morphology of the emission-line region appears to be directly 
related to that of the radio emission. In particular, in Seyferts with radio 
jets (e.g. Mrk 3, Mrk 348, Mrk 6, Mrk
1066, ES0 428-G14), the NLR itself appears jet-like and is
spatially coincident with the radio jet, while
the emission-line region takes a different form when a radio lobe is
present (e.g. Mrk 573, Mrk 78, NGC 3393): 
arc-like shells of emission, very reminiscent of bow-shocks, 
surround  the leading edge of the lobes 
(Capetti et al. 1995a, 1995b, 1996; Falcke et al. 1996, 1998).
This dichotomy in radio and
emission-line morphology is reflected in their different scales: 
bow-shock structures cover several kiloparsecs,
while the jet-like features extend only over a few hundred parsecs.  
The simplest interpretation of this radio-to-optical correspondence is 
that the radio emitting outflow creates an expanding and cooling gas halo.
The compression induced by the outflow causes the line emission to be 
highly enhanced in the regions where the jet-cloud interactions occur.
A clear confirmation of this scenario came recently from HST spectroscopy 
of Mrk 3 (Capetti et al. 1999): its NLR has velocity field characteristic 
of a cylindrical shell expanding at a rate of 1700 km/s. They
interpreted this as the consequence of the rapid expansion of a hot gas cocoon 
surrounding the radio-jet, which 
compresses and accelerates the ambient gas. 

HST observations also provided
evidence for spatial variations in the NLR ionization structure.
In NGC 1068 the material located along the radio jet is in a much higher 
ionization state than its surroundings. This might suggest the presence of a 
local source of ionization which dominates over the 
nuclear radiation field (Capetti, Axon and Macchetto 1997; Axon et al. 1998). 
In other sources, too distant for such a detailed analysis, 
the radial variations of the ionization parameter are generally much
flatter than expected from pure nuclear photoionization on the basis of
the measured density gradients (Capetti et al. 1996, Allen et al. 1999) 
requiring again a local source of ionizing photons. 
An appealing possibility of interpreting these data is
to invoke the ionizing effects of 
shocks, originated by  jet-cloud interactions: 
if these shocks are  fast enough (velocities $>$ a few 
hundred km s$^{-1}$) the hot, shocked gas could produce a significant flux 
of ionizing photons (Sutherland, Bicknell and Dopita 1993; 
Dopita and Sutherland 1995, 1996). Direct evidence for this emission 
has been found by
Axon, Capetti and Macchetto (1999) who showed that the radio-jets 
in the Seyfert 2
galaxies Mrk 348 and Mrk 3 are associated with an extended linear
structure in UV and optical continua. 
In this picture, the radio-jet would not only 
determine the morphology of the NLR but is physically involved in
its ionization. 
A radio imaging survey of the CfA sample of Seyfert 
(Kukula et al. 1995) shows that
radio linear structures are present in a large fraction of sources (more than 
50\%) suggesting that such an interaction is likely to be a quite common
phenomenon in this class of objects.  

The jet interaction with the external medium is clearly 
a complex physical problem which involves both a hydrodynamical study 
of the jet propagation as well as a detailed understanding of the 
microphysics of the induced shocks, which might also be magnetized, and of the 
radiative processes.

In the framework of Seyfert galaxies this issue has been tackled by several
authors (Dopita and Sutherland 1995, 1996,
Evans et al. 1999, Wilson and Raymond 1999, Allen et al. 1999).
Their focus is however mainly on the shocks properties with a very 
detailed treatment of the emission mechanisms, with simplifying assumptions
about the hydrodynamics (e.g. plane parallel 
geometry, steady-state shock).
The comparison with the observations is 
based on the emitted spectrum and in particular on 
diagnostic line-ratios, particularly with the aim of distinguishing the 
different signatures of nuclear versus local photoionization. 

In this paper we follow a complementary,
albeit different, approach by studying in detail the 
jet hydrodynamics, while adopting a simplified treatment of the radiative 
processes, as we employ an equilibrium cooling 
function in an optically thin approximation. 
This approach allows us to compare
the results of simulations with
the observed properties of NLR, in particular their morphology, 
the expansion velocities and the characteristic 
values of gas density and temperature. 
More precisely we consider the interaction
of the jet with an inhomogeneity in the external medium (cloud) and 
 our aim is that of constraining
the jet and cloud physical parameters for which it is possible to 
reproduce the observed conditions.
In this way, in addition of getting a better understanding of the  
NLR physics, we can also obtain information on the 
jet properties from the NLR data. Moreover, we can  calculate the fraction of
the jet power converted in radiation by shocks, resulting from the
interaction of the jet with the environment. We then get from the global 
dynamics a conversion efficiency from kinetic to radiative power
 and we can determine whether
the jet itself, via shocks, can provide an {\it in situ}  
photoionization source  for the NLR emitting material,
as discussed above. 

Steffen et al. (1997a) have used a rather similar approach 
with the main difference that they considered the jet 
propagating into a uniform medium. It seems that in this 
situation it is impossible to reach the high densities typical of the 
NLR with jet-like emission (see discussion below) 
on which we will focus in the present
paper. This is because, at low density, radiation is not efficient enough to
give the needed compression factors.
The conditions of the emitting material obtained 
by Steffen et al. seem to be appropriated for the case of the more 
extended (lobe-like) line emission structure. 

Steffen et al. (1997b) considered also jet-cloud interactions
mainly from an analytical point of view. They found that
when a jet interacts with a large number of clouds the most relevant effects
on the NLR structure
are due to the most massive clouds located along the jet path.
This lead us to our choice for the geometry of the simulation in which
the jet hits a single dense cloud.

The paper is structured as follows: In the next section (Sect. 2), we 
describe the basic physical problem and the observational constraints,  
while the equations used and the method 
of solution are examined in Sect. 3 and 4; the results of simulations are 
discussed in Sect. 5; conclusions are drawn in Section 6. 
\section{Observational data and astrophysical scenario}
Observational data provide us with quite detailed information on the physical
conditions of the narrow line emitting regions, in particular HST observations
can now be used to determine the propertirties of individual NLR clouds:
 typically, densities are 
larger than $10^3\,$cm$^{-3}$,  temperatures are of the order 
of $10^4 - 2\times10^4\,$ K, and 
velocities are $\sim 300 - 1000\, {\rm km \ s}^{-1}$ (Caganoff et al. 1991,
Kraemer, Ruiz and Crenshaw 1998, Ferruit et al. 1999, Axon et al. 1998,
Capetti et al. 1999).   

These are the observational constraints that we try to match 
in our simulations.
Results of simulations of a jet impinging on a uniform medium,
with properties typical of the ISM,
have shown that it is not possible to match, in this situation, the density
values reported above (Steffen et al. 1997a, Rossi \& Capetti 1998). 
We will therefore consider throughout the rest of the  
paper the case of a jet impinging on pre-existing inhomegeneities. 
We can identify such inhomogeneities  with giant molecular clouds (GMCs), 
that typically populate spiral galaxies.
These objects have typically mass $\sim 10^5 - 10^6 \ {\rm M}_{\odot}$, radius
$\lapp 100$ pc, and temperature $\sim 10$ K (Blitz 1993). 
The resulting particle densities
span from a few up to about hundred particles per cm$^3$. 

A supersonic
jet, of radius $\sim 10$ pc, that bore its way through the interstellar medium
has a considerably good chance of impinging frontally upon a (much larger) GMC,
and this is the case we will consider in our simulations. In any event, 
this latter case, i.e. the head-on collision with a large cloud, can be considered 
the most efficient 
case of interaction, for the compression, acceleration and heating
of the NLR material.

As discussed below the effects of the jet/cloud interaction last for a time 
considerably longer than the cloud crossing time.
Moreover, the jet crosses the tenuous inter-cloud regions at a much higher 
speed than while in a cloud. We therefore expect that more 
than one cloud will be interacting at any given time
and they will display simultaneously the different evolutionary stages
of the interaction.

\section{ The physical problem} 
We study the evolution of a cylindrical fluid jet impinging upon a cold heavy 
steady inhomogeneity, namely the cloud, in pressure equilibrium with the 
external medium. The relevant equations governing the jet evolution, 
for mass, momentum conservation, and  radiative losses, are 
$$ 
{{\partial \rho } \over {\partial t}} + \nabla \cdot (\rho \vec{v}) = 0 
\,, 
\eqno (1{\rm a}) 
$$ 
$$ 
{{\partial \rho v_r} \over {\partial t}} + \nabla \cdot { 
(\rho v_r \vec{v })} = - {{\partial p}\over {\partial r}}\,, 
\eqno (1{\rm b}) 
$$ 
$$ 
{{\partial \rho v_z} \over {\partial t}} + \nabla \cdot { 
(\rho v_z \vec{v })} = - {{\partial p}\over {\partial z}}\,, 
\eqno (1{\rm c}) 
$$ 
$$ 
{{\partial E} \over {\partial t}} + {\nabla \cdot (E \vec{v })} = 
- p \nabla \cdot \vec{v } - {\cal L}\,,
\eqno (1{\rm d}) 
$$ 
where the fluid variables $p$, $\rho$, $\vec{v}$ and $E$
are, as customary, pressure, density, velocity, and 
thermal energy ($p / (\Gamma -1)$) respectively; 
$\Gamma$ is the ratio of the specific heats; ${\cal L}$ represents  
the radiative energy loss term 
(energy  per unit volume per unit time, 
Raymond and Smith 1977).

The jet  occupies  initially   
a cylinder of length $L$. 
The initial flow structure  has  the following form:
$$
v_z(r) = 
\left\{ \begin{array}{ll}
    {v_z(r=0)}\over{\cosh[(r/a)^m]}   & \qquad {z \le L} \\
    &\\
    0 & \qquad {z > L}
\end{array} \right.
\nonumber
$$
where $m$ is 
a `steepness' parameter for the shear layer separating the jet from the
external medium (see Fig. 1).
The choice of separating the jet's interior from the ambient medium with a 
smooth 
transition, instead of a sharp discontinuity,  avoids numerical instabilities
that can develop at the interface between the
jet and the exteriors, especially at high Mach numbers.

Regarding the cloud, we fix its initial density $\rho_{\rm cloud}$ and impose 
pressure equilibrium with respect to external medium; for simplicity we 
consider 
a steady cloud, with a thickness equal to the jet diameter.
 \begin {figure}
 \psfig{file=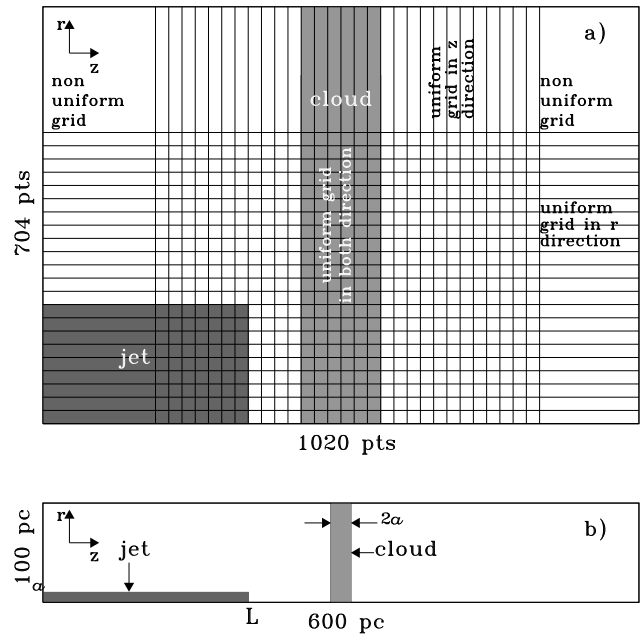,width=\hsize}
 \caption []
 {In panel a)  the computational domain is sketched. The grid is finer on the
 region of jet/cloud interaction, while is coarser far from the region of our
 interest. In panel b) the physical domain is shown.
 }
 \end{figure}
\section{ The numerical scheme } 

\subsection{Integration domain and boundary conditions}
Integration is performed in cylindrical geometry and
the domain of integration ($0 \leq z \leq D$,  $0
\leq r \leq R$) is covered by a grid of $ 1020 \times 704$ grid points.
The axis of the beam is taken coincident with the bottom boundary
of the domain ($r=0$), where symmetric (for $p$, $\rho$ and $v_z$) or
antisymmetric (for $v_r$) boundary conditions are assumed. At the top
boundary ($r=R$) and right
boundary ($z=D$) we choose free outflow conditions, imposing
for every variable $Q$ null gradient ($dQ/d(r,z)=0$). 
The boundaries are  placed
as far as possible from the region of the jet where the most interesting
evolutionary effects presumably take place by employing
a nonuniform grid both in the longitudinal $(z)$ and the radial $(r)$ 
directions
(Fig. 1, panel a)).
In the radial direction the grid is uniform over the first 500 points and
then 
the
mesh size is increased assuming $\Delta r_{j+1} = 1.015 \Delta r_{j}$. The
jet spans over 200 uniform meshes, while the external boundary is
shifted to  $r = 10a$ where $a$ is the jet radius.
As for the $z$--direction, we assume a constant fine grid in the
central part of the domain, where the cloud is located, 
i.e. in a sub-domain of length $40 a$, between the grid points 180 and 
844; conversely, in the remaining part we consider an
expanded  grid increasing the mesh
distance according to the scaling law $\Delta z_{j \pm 1} = 1.015
\Delta z_{j}$,
where the minus sign applies in the first 180 grid points and the plus sign
above grid point 844.  

\subsection{ Integration method } 

 The basic equations (1a-d) have been integrated with a two-dimensional 
version of the Piecewise Parabolic Method (PPM) of Colella \& Woodward 
(1984) (for a discussion of the main characteristics of this code 
 and its merits for this kind of problems see Bodo et al. 1995).
 Radiative losses are dealt with the operator splitting  
 technique, following which we split a single time step into two parts. 
 In the first part, we advance the dynamical quantities,  
 by using the adiabatic equations.  
In the second part we update the internal energy,  
keeping all the other variables constant, by taking into account  
radiative losses.  

%
\subsection{ Physical parameters and Scaling} 
The physical problem that we are approaching is quite complex, 
with three different 
interacting and radiating media, i.e. jet, ambient medium
and cloud, each one described by its density, 
temperature, velocity and size.
We note that in the adiabatic simulations of propagating jets, 
by normalizing to the jet density, sound speed, jet radius and
 sound crossing time over the jet radius, we are
left with only two parameters, namely the density ratio between jet and 
external medium and the jet Mach number.
The presence of radiation complicates the matter (Rossi et al. 1997), in fact 
temperature is not  scale free, since the radiative loss function in 
Eq. (1c) explicitly depends on its physical value and in addition to the 
sound crossing time ($t_{\rm cr} = a/c_{\rm s}$), we have another typical 
 time scale of the system, i.e. the radiative  
 time scale, defined as $t_{\rm rad} = p/[(\Gamma-1){\cal L} ] $ which depends
on the density of the medium.  Therefore in this case one has to consider
for each medium the value of density and temperature as independent parameters.
In addition, as we already noticed we are now considering three media.  
When the jet passes through the cloud,  
the evolution of the compressed cloud material is completely 
different with respect to the case of  two media,
where the jet continues to push  
dense material at the head and it does not
have any reaccelerations related to 
the passage from a denser to lighter medium. We would like to stress that the 
presence of a inhomogeneity is fundamental, in fact only in this 
case, as we show 
later, it is possible to reach the proper density for the emitting material.
In conclusion, we must assign a large number of physical parameters for
defining the initial conditions of our simulations.

A thorough investigation of such huge parameter space is unfeasible; however,
 not all the parameters are equally important and some of them can 
be well constrained by observational considerations. As a first step  
we will then fix criteria  to minimize the number 
of free parameters.

Concerning the external (uniform) medium, we have to fix $\rho_{\rm ext}$ and
$T_{\rm ext}$,  having $v_{\rm ext} = 0$.  With respect to  
$\rho_{\rm ext}$, we can assume one particle per cubic centimeter, 
a value which we know to be appropriate to the interstellar medium of our 
Galaxy (Cox \& Reynolds, 1987). 
In reference to $T_{\rm ext}$, again its choice is not so crucial,
since  the most 
important temperature for the emission processes is the shock temperature, 
depending mainly on jet velocity, in any case  observations tell us that 
the external medium is completely ionized, which means temperatures
larger than $10^4$ K, and we assume $T_{\rm ext} = 10^4$K. 

The jet is physically described 
by its density  $\rho_{\rm jet}$, temperature $T_{\rm jet}$, 
 initial velocity $v_{\rm jet}$ and  radius $a$. 
The radius $a$ can be chosen as our  length unit in order 
to scale the other lengths in the system, and following the radio
observational suggestions (Pedlar et al. 1993, Kukula et al. 1999) 
we consider it to be $10$ pc. 
Concerning the jet density we do not have any tightening constraint, so in a 
first approach, we take it equal to $\rho_{\rm ext}$.
Relatively to  $T_{\rm jet}$,  looking at the loss function (Fig. 2) 
we can immediately realize that its 
initial value it is not so crucial, since cooling is fast 
and  soon the jet temperature falls to $\sim 10^4$ K. Anyway $T_{\rm jet}$ in
our simulations is taken to be $10^6$K.
The jet velocity will be instead an important parameter of our simulations.

Finally we consider the cloud, its density $\rho_{\rm cloud}$ is the
parameter on which 
we will focus our investigation. $T_{\rm cloud}$ will be fixed by imposing 
pressure equilibrium with the
external medium. Actually GMC's are not required to be in pressure equilibrium
since they might be autogravitating, however, as discussed for the external 
temperature, the exact value of  $T_{\rm cloud}$  is not crucial for the
results of the simulations. For simplicity we consider a steady cloud
($v_{\rm cloud} = 0$). The cloud dimensions must lie in the range of
GMCs, so we will fix the size longitudinal to the jet to 20 pc
(i.e. $2a$), with
a indefinitely large (with respect to $a$) transversal size.

In summary, we have three control parameters, namely the initial cloud 
density, the initial 
jet velocity and  the initial jet density, that we fix equal to one
particle per cm$^{-3}$. So we will 
investigate in details the effects of adopting different values
for $v_{\rm jet}$ and $\rho_{\rm cloud}$.
\begin {figure}
 \psfig{file=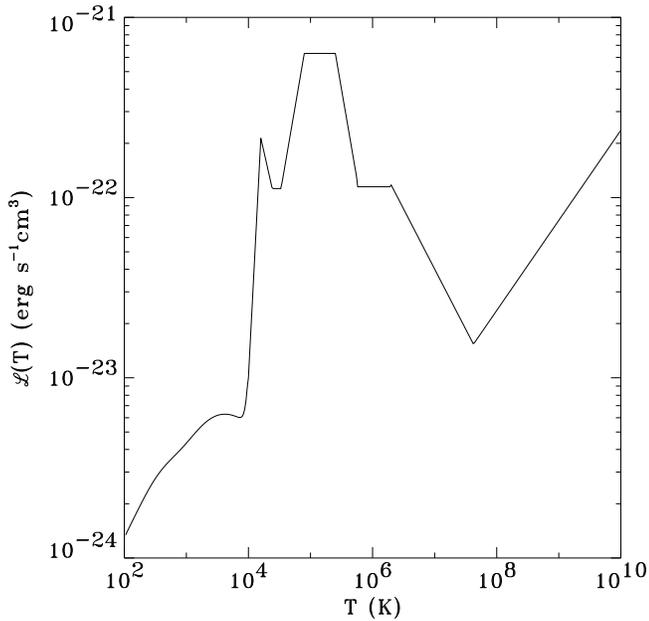,width=\hsize}
 \caption []
 {Plot of the energy loss function {\sl vs} temperature 
(Raymond \& Smith 1977).}
 \end{figure}
\section{ Results} 

We begin
our discussion with a short general description of the complete evolution 
of the jet-cloud interaction, that can be summarized in three steps
(see Fig. 3 for a visualization of the basic features of the three steps for
the case $\rho_{\rm cloud}=30\, {\rm cm}^{-3}$ and $v_{\rm jet}=6500\,
{\rm km \ s}^{-1}$):
 \begin {figure*}
 \psfig{file=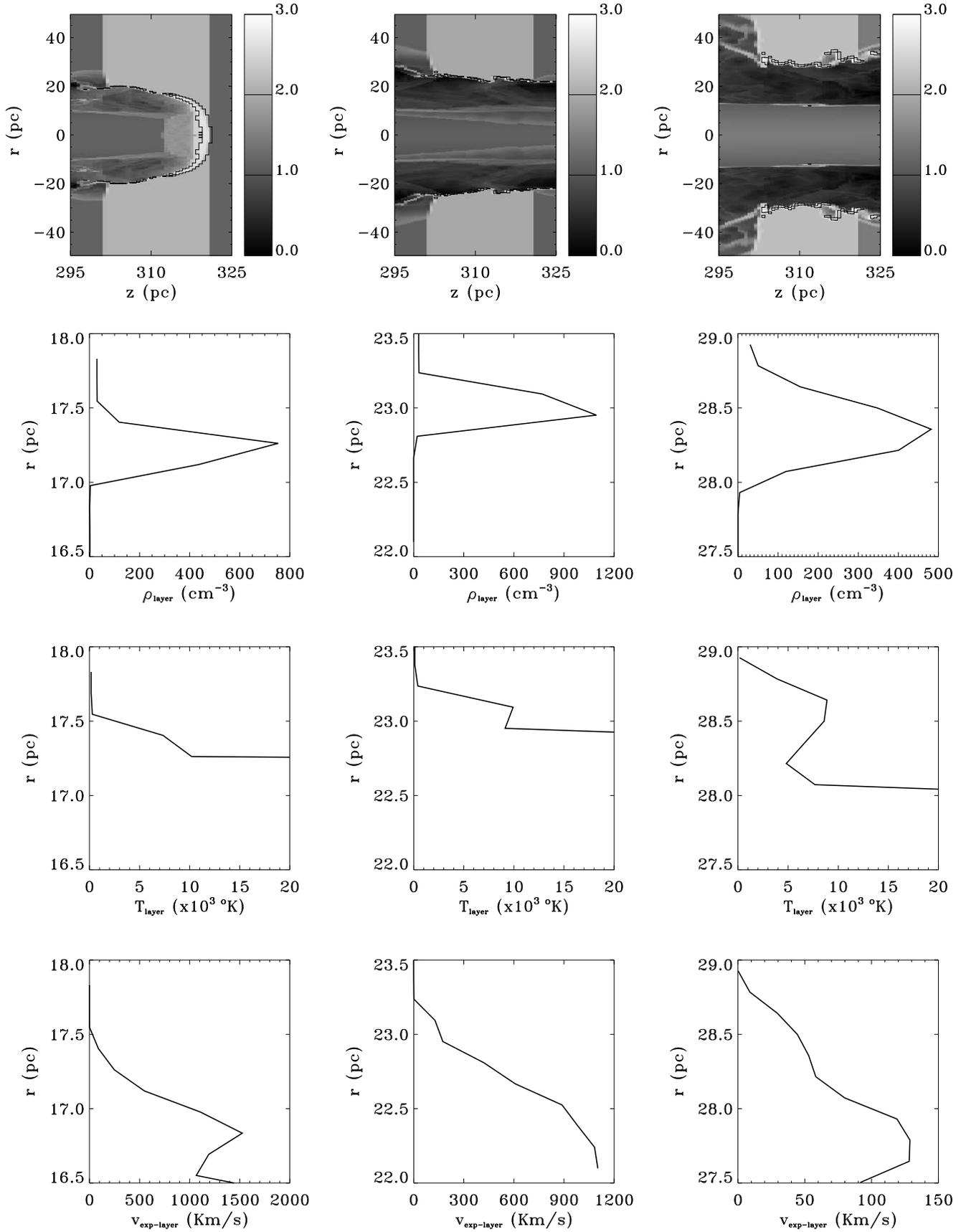,width=\hsize}
 \caption []
 {Images of the density distribution showing the jet-cloud interaction 
and cuts of density, temperature and
expansion velocity, in the middle of the cloud, across the thin  layer 
of compressed material for the case $\rho_{\rm cloud} =
30\,{\rm cm}^{-3}$, $v_{\rm jet} = 
6500\,{\rm km \ s}^{-1}$, at different times. 
The three columns correspond respectively to $t=1\,t_{\rm cc}$,
$t=2\,t_{\rm cc}$ and 
   $t=5\,t_{\rm cc}$.}
 \end{figure*}

\noindent
{$\bullet$} The jet hits the cloud, forming a strong shock, the post-shock 
region becomes hot and blows up, because of its increasing pressure; the jet 
material is conveyed in a back-flow that squeezes the jet itself. During this 
process the cloud material is compressed and heated by the shock, at the head 
the temperature is very high ($> 10^8$ K), while on the jet sides it is lower 
$\sim 10^7$ K, so that it can cool down, to reach the observed line emission 
conditions. It is in this region, forming a layer around the jet,  that the 
narrow line emission can originate. Our analysis 
will therefore concentrate on the  properties of this region. 
During this first phase, in which the jet 
crosses the cloud, the layer is accelerated by the strong inside pressure
and cools down,  its density thereby increases 
(see the leftmost panels in Fig. 3).

\noindent
{$\bullet$} The second phase begins when the jet is completely out of the
cloud, 
the compressed emitting material reaches a quasi-steady state, during which 
the emission is almost constant, the inside pressure begins to decay, but the
 emitting layer is still accelerated. From Fig. 3 (central panels), we 
can see that the material in the layer has been compressed, its maximum 
density has increased, while its temperature has decreased. The maximum
density is found now at temperatures around $10^4$ K, and its velocity 
has also increased. 

\noindent
{$\bullet$} In the third phase, the inside pressure has decayed and the 
emitting layer begins to slow down, the jet flows freely through the cloud and 
also the emission decreases, eventually disappearing. 
From Fig. 3 (rightmost panels),
we see a decrease in density and velocity, while almost all the layer is found 
at $5 \times 10^3 < T < 10^4$ K.

The efficient formation of the line emitting region will therefore depend on 
the efficiency of radiation during the jet crossing of the 
cloud. We will then 
introduce two typical timescales, the 
{\it cloud crossing time} and the {\it radiative time},
whose ratio will be a fundamental parameter for determining the evolution 
of the narrow line emitting layer. 
Following analytical treatments of the jet-ambient 
interaction  we 
define the cloud crossing time as 
$t_{\rm cc} = d ( 1 +\sqrt {\rho_{\rm cloud}/\rho_{\rm jet}})/{v_{\rm jet}} $, 
where we have assumed, for the jet head velocity in the cloud,
the steady velocity obtained from the 1-D momentum balance in a medium 
with $\rho = \rho_{\rm cloud}$ (see, e.g., Cioffi and Blondin 1992,
 Norman et al.\ 1982). In this way we are actually overestimating
the crossing time, since our situation is not steady, however this value
is sufficiently accurate for our purposes.
Concerning the radiative time, its definition is given in Section 4.3,
however, we must notice 
that for  its evaluation we have to assume a value for the temperature,
in the following considerations we have taken $T = 10^7$ K, that 
is the average of the typical post-shock 
temperature in the region of our interest, this choice is properly done for 
all jets with $v_{\rm jet} = 6500 \ {\rm km \ s}^{-1}$ and  
$v_{\rm jet} = 32500 \ {\rm km \ s}^{-1}$, 
while it is overestimated for the low velocity cases, that means that  
$t_{\rm rad}$ for those cases are shorter than the real ones. We have
then defined
$\tau \equiv t_{\rm cc} / t_{\rm rad}$ as the ratio between crossing
and radiative 
time scales and this, as said before, is an important parameter for the 
interpretation of the results.

As a first step in our analysis, we have performed an exploration of the
parameter space.           
As discussed before, we reduced our parameters to $\rho_{\rm cloud}$ and the 
initial $v_{\rm jet}$. In Table 1 we report, for each pair
of their values, typical values of density, 
expansion velocity and temperature of the emitting material
and the value of $\tau$. 
The density is the median value of density distribution weighted on the 
emissivity function (that is proportional to $\rho^2$), while velocity 
and temperature are those corresponding to this density value. 
All the quantities are evaluated at $2\,t_{\rm cc}$, this choice is due to 
the fact that during this period the expansion velocity of the emitting 
material increases rapidly 
reaching a maximum and then decreases monotonically, so that, if the
expansion 
velocity does not match the observational constraint within  this time,
it never will, and the case will not be of interest for our analysis.
Radiation must therefore act efficiently during this time, in order to 
create the needed conditions for radiation, and this poses a lower limit 
on the value of $\tau$. On the basis of the values reported in this table 
we choose  the most promising cases for our investigation. 

\noindent
\begin{table}
\caption{Parameters of the simulations}

\input tabella.tex

\end{table}

Considering the first column we can immediately realize that jets at low 
velocity cannot reach conditions comparable to those observed. 
The values of $\tau$ for these simulations are high, meaning that radiation 
is very efficient. On the other hand,  the jet momentum is low and 
cannot drive the emitting material at high velocities. For the case 
$\rho_{\rm cloud} = 30 \ {\rm cm}^{-3}$ we have, in fact, high densities in 
accord with the high 
value of $\tau$, but very low velocities. For this reason we did not   
perform simulations for the other two cases
of higher density, since jets 
would produce  stronger and cooler compression practically at 
rest, very far from the observational scenario. 

Looking at the high velocity case, we see that, in the case of small 
inhomogeneities, $\tau$ has a very low value and, therefore, radiation is 
inefficient. The jet is very energetic and sweeps the 
cloud, before radiation becomes effective and so it does not form 
any condensation (the velocity reported for this case is therefore 
meaningless). Increasing the cloud density, we increase also the value of 
$\tau$: the maximum density increases, but it is still quite low. Only for 
the high density cloud ($\tau = 0.5$), we get values of density and velocity
in agreement with observations.
 
Regarding the intermediate velocity, the values of $\tau$ are $> 0.3$: 
 radiation is efficient and thus the emitting layer can reach sufficiently
high densities. Only in the lighter 
cloud case, however, the velocity is comparable to the observed values.

From this exploration of the parameter space we can conclude that the
observed 
conditions can be matched only for a narrow range of parameters and that 
the properties of the emitting layer depend 
essentially only on one parameter, the ratio between the 
radiative timescale and the cloud crossing timescale $\tau$.
For low values of $\tau$ ($\tau < 0.3$), radiation is inefficient and the 
densities in the layer are too low. For higher values of $\tau$ 
($\tau > 0.55$) we find, on the other hand, that the velocity of the emitting
layer becomes too small. This is because the cloud density is high and the jet 
momentum flux is too small to impart to it a large enough velocity. 
Only for a narrow range of values of $\tau$ 
we can match the observed conditions and, in Table 2, we have
translated these limits into limits on velocity range at different cloud
densities. 

\begin{table}
\caption{Ranges of jet velocities that can match the observed properties}

\input tabella1.tex

\end{table}
\noindent
\subsection{Case of $\rho_{\rm cloud}\,=\,120\ {\rm cm}^{-1}\,,\,v_{\rm jet}\,=
\,32500\ {\rm km \ s}^{-1}$ 
} 
In this subsection we will discuss in more details the case that 
best matches the 
observational scenario.  We begin our discussion showing, Fig. 4, 
a gray-scale image with a snapshot of 
the density distribution at $5t_{\rm cc}$  and  
three small panels showing enlargements of the region of interaction 
between jet and 
cloud   referred to  density, temperature and the expansion velocity of 
emitting gas. The proper physical condition  for emission are 
reached  in 
a thin layer of compressed cloud material, whose width and mass grow 
in time as 
the shocked cloud material cools down.

\begin {figure*}
\psfig{file=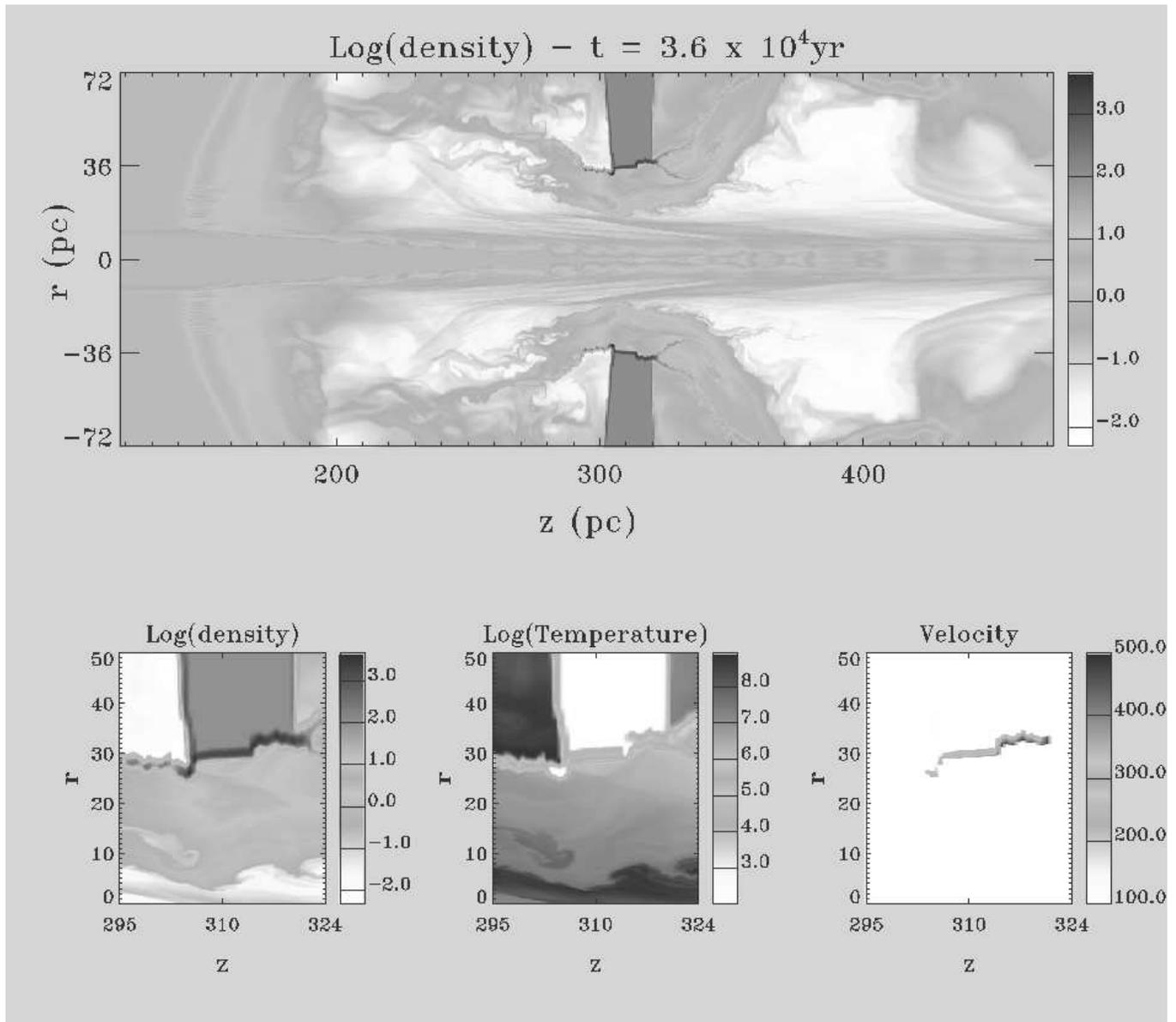,width=\hsize}
\caption []
{The larger panel shows the image of the distribution of logarithm 
of density at 
$5t_{cc}$ for the case $\rho_{\rm cloud} =
120\,{\rm cm}^{-3}$, $v_{\rm jet} = 
32500\,{\rm km \ s}^{-1}$. The other three small panels show, in enlargements
 of the interaction region,  the distributions of the logarithm of density, 
the logarithm of temperature and of velocity.}
\end{figure*}

   The detailed physical properties of this line emitting region
are reported in Fig. 5, where we have represented the 
behavior of density, temperature and velocity along
radial cuts through this layer. We note that the proper conditions
are matched in a layer of width $\lapp \,2$ pc.

\begin {figure}
\psfig{file=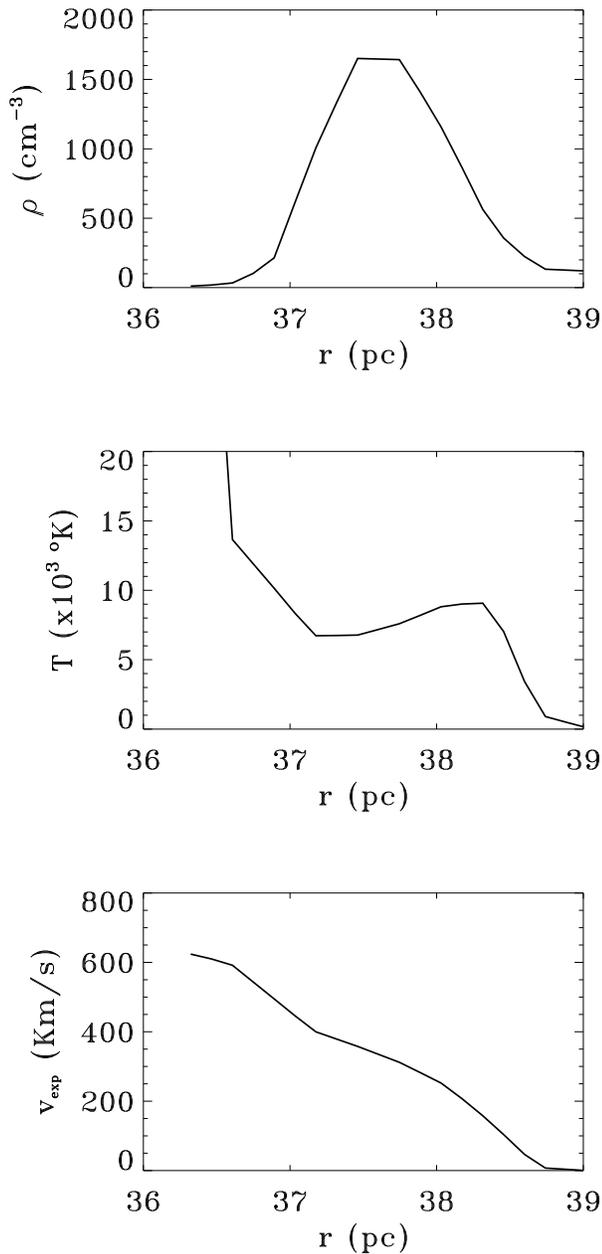,width=\hsize}
\caption []
{The three panels present a transversal cut of 
 density, temperature, and the expansion velocity, in the middle of cloud,
across the thin emitting layer for the same case of Fig. 4. Notice that the 
coordinate $r$ measure the distance from the jet axis.}
\end {figure} 
 
How the properties of the material contained in this thin layer 
compare with the 
physical conditions of gas of the NLR? 
To answer the question we plot in
Fig. 6  the temporal behavior  of the mean 
expansion velocity (panel a) and mass (panel b) of the emitting 
material shell at two different density limits. We see that from the time
when the jet touches the cloud until 
$2t_{\rm cc}$, when 
 a strong interaction between the jet head and the cloud takes place, 
the cloud material is accelerated and 
the quantity of emitting material increases; 
after this  interval the jet flows, essentially freely, across the 
cloud without any further acceleration of the compressed material 
shell and the accelerated cloud material  slows down  monotonically.
\begin {figure} 
\psfig{file=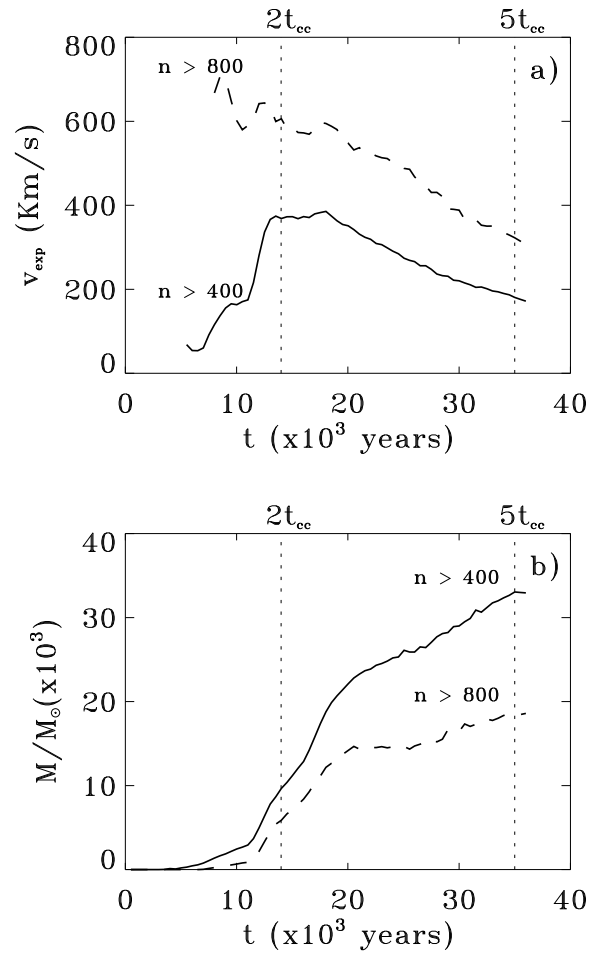,width=\hsize}
\caption []
{In the two panels we plot the behavior of the mean expansion velocity of 
the emitting material vs 
time (panel a) and of the total emitting mass, in unit of solar masses, 
vs time (panel b). The solid 
lines refer  to   material denser than 400 particle per cube centimeter,while 
the dashed line refers to material denser than 800 particle per cube 
centimeter.}
\end{figure}
Notice that the mean expansion velocity, relative to an observer, lies,
for the denser material, in the range
 $600-1200\ {\rm km \ s}^{-1}$  
(since one must consider twice the mean expansion velocity), that is in 
good agreement with the velocity deduced by the line widths detected.
Looking more in detail at the emitting mass,
 we see that its growth begins some time after the jet has initiated 
to drill its way
 into the cloud, and this delay corresponds to the cooling time of the 
shocked material. We also note that, after $t=2t_{\rm cc}$, the jet 
continues to
sweep out material laterally at a pace that is higher for the lighter
material, the total mass exceeds
$3\times 10^4 / {\rm  M}_{\odot}$ at $t=35,000$ ys and this would correspond 
to an $H_\beta$ luminosity of $\sim 2 \times 10^{40}$erg s$^{-1}$ which,
considering also the possibility of having simultaneuously several active 
clouds, is consistent with the observed values.

As the interaction is effective over a timescale much longer than 
$t=t_{\rm cc}$ the jet will quickly propagate into the low density inter-cloud
medium and it will reach any other cloud lying on its path. 
Thus more than one cloud
will be effectively interacting with the jet at any time. Each will display
a behaviour typical of its evolutionary stage and the total emitting mass
must be considered as the total over all clouds. Furthermore, this will 
naturally reproduce the jet-like morphology of the NLR.

\subsection{Energetics} 
\begin{table*}
\caption{High frequency radiative power and radiative efficiencies}

\input tabella2.tex

\end{table*}

As discussed in the Introduction, the source of ionization of the NLR 
is still matter of debate. While the NLR gas is certainly illuminated
by the nuclear source, its interaction with the radio jet also produces
regions of high temperature and density which radiates ionizing photons.
In this paragraph we derive the conversion efficiency of the jet kinetic
power into energy radiated in ionizing photons. To estimate the 
ionizing energy flux we integrated radiative losses over all regions where 
$T > 10^5$ K as above this temperature most of them 
correspond to the production of photons with energy higher than the
hydrogen ionization threshold.

In Table 3 we summarize  our results  reporting the 
kinetic power referred to the three different velocities
($P_{\rm kin}=\rho \ v_{\rm jet}^3 \ A$,
where  $\rho$, $v_{\rm jet}$ and $A$ are respectively the density,
the velocity and the 
transverse section of the jet) and the conversion 
efficiency at peak and after 2 $t_{cc}$ for all the cases considered.

The peak efficiency reaches in one case a value as high as 10 \%
but it is usually 0.1 - 2  \%. 
However, over the interaction, the typical value of $\eta$ 
(well represented by its value after  2 $t_{cc}$) is much lower 
$\eta \sim  10^{-4} - 5. \times 10^{-3}$. Faster jets have lower efficiency 
than slower jets and this conspires in producing a very similar  
amount of energy radiated in ionizing photons, 
$\sim 10^{40} \ {\rm erg \ s}^{-1}$, in all cases.
 
In Seyfert galaxies $L_{H\beta} \sim 10^{39} - 10^{42}
{\rm erg \ s}^{-1}$ (Koski 1978).
The minimum ionizing photon luminosity required to 
produce a given line emission luminosity corresponds to the limiting case in 
which all ionizing photons are absorbed
and all photons have an energy
very close to the hydrogen ionization threshold $\nu_{\rm ion}$.
In this situation

$  L_{\rm ion,min} = \frac{1}{p_{H\beta}}\frac{\nu_{\rm ion}}{\nu_{H\beta}} 
 L_{H\beta} \sim 50 \, L_{H\beta} $

\noindent
where $p_{H\beta}\approx 0.1$ is the probability that any recombination will 
result in the emission of an H$\beta$ photon.

It appears that, even in this most favourable scenario,
the radiation produced in shocks 
can only represent a small fraction
of the overall ionization budget of the NLR, particularly as sources with
high radio luminosity (in which usually radio-jets are found)
also have the highest line luminosity (e.g. Whittle 1985).

Nonetheless, in the most promising case examined above
($\rho_{\rm cloud}=120\, {\rm cm}^{-3}$ and $v_{\rm jet}=32500\,
{\rm km \ s}^{-1}$)
at the peak of the conversion efficiency the radiated energy 
is $3 \times 10^{41} \, {\rm erg \ s}^{-1}$ and it is substained over 
a crossing time, $\sim 10^4$ years. Shock ionization may thus produce
important ionization effects which, however, can be only
both local and transient. 

\section{Conclusions} 

We have studied in detail the dynamics of the 
interaction of a jet with a large cloud 
pre-existing in the ISM in order to find the conditions for which it
is possible to reproduce the main physical parameters of the NLR emitting
material. 

Following the suggestion by Steffen et al. (1997b) 
that the most relevant effects of the interaction arise when a jet hits
dense massive clouds, we adopted a quite simplified geometry of a single 
gas condensation with can be astrophysically identified with a giant 
molecular cloud. 
As the interaction last for a time 
considerably longer than the cloud crossing time 
more than one cloud will be interacting at any given time
and they will display simultaneously the different evolutionary stages
of the interaction. Furthermore the characteristic jet-line structure 
of the NLR is thus reproduced. In any event, 
this case, i.e. the head-on collision with a large cloud, is 
the most efficient case of interaction, for the compression, acceleration 
and heating of the NLR material. 

We concentrated  our efforts on the exploration of
the parameter plane ($v_{\rm jet}, \ \rho_{\rm cloud})$, since the other 
parameters, on which the simulation depends, have little influence on
the properties of the optically emitting material. 
We have found that the condition
for obtaining values of density, temperature and velocity in the observed
range can be translated in a condition on the parameter  $\tau$, 
which is the ratio of the cloud crossing timescale to the radiative timescale
($0.3 < \tau < 0.55$) and which depends on our two fundamental parameters
$v_{\rm jet}$ and  $\rho_{\rm cloud}$. For small values of $\tau$, radiation
is inefficient and it is not possible to produce regions dense enough, while,
on the other hand, for large values of $\tau$, the cloud is too dense 
and the obtained
velocities are too low. We have explored a range of cloud densities which 
can be considered typical of GMCs and, for this range,
the jet velocities  span an interval
from  $4000\,$km s$^{-1}$ to $55,000\,$km s$^{-1}$.

The jet kinetic power corresponding to these combinations of parameters 
(for a jet density of 1 cm$^{-3}$) ranges from
$3.2 \times 10^{41}$ to $8 \times 10^{44}$ erg s$^{-1}$, in general agreement
with the estimates of Capetti et al. 1999 for Mrk 3. 
For jet density much lower than 1 cm $^{-3}$, however, in order to
match the observed NLR conditions we would need a correspondingly higher
velocity and therefore untenable requirements on the kinetic power
which grows with $v_{\rm jet}^3$. 
We conclude that jets in Seyfert galaxies are unlikely to have densities
much lower than 1 cm $^{-3}$ and velocities higher than  
$50,000\,$km s$^{-1}$, and therefore they are very different from 
their counterparts in radio-galaxies in which densities are 
much lower and velocities are relativistic. 

Concerning radio-galaxies we can speculate
that with lower jet densities and higher velocities,
the gas postshock temperatures and radiative time 
would be increased with respect to the case of Seyfert galaxies
and therefore the conditions for having efficient 
line emission would be more difficult to meet. 
In addition the different properties of the jet environment 
in the elliptical galaxies hosting radio-galaxies render
encounters with gas condensations less likely to occur. 
This probably explain why 
the association between radio and line emission although 
often present in radio-galaxies (e.g. Baum and Heckman 1989)
is not as strong as in Seyfert galaxies.

Finally, the study of the global dynamics allowed us to have 
estimates of the overall efficiency of the conversion of kinetic
to high frequency radiative power in the shocks that form in the interaction 
between jet and ambient medium. We have found that the efficiency is 
increased by the presence of the cloud, its peak value is 
0.1 - 2 \% , its typical value is much lower $\sim 10^{-4} - 5 \times 10^{-3}$
and it decreases with the jet power. These results lead us to the conclusion
that radiation emitted in shocks can be only a small fraction of the
overall ionization budget of the NLR, although it can have local and 
transient important effects. 
 
\begin{acknowledgements} 
We thank CNAA (Consorzio Nazionale per l'Astronomia e Astrofisica) for
supporting the use of supercomputers at CINECA, .
\end{acknowledgements}

\end{document}

%% file: tabella.tex
\vspace{0.30cm}
{\centering \begin{tabular}{|c|c|c|c|}
\hline 
&&&\\
&$v_{jet} =$&$v_{jet} =$&$v_{jet} =$\\
&&&\\
&$1300\,km\,s^{-1}$&$6500\,km\,s^{-1}$&$32500\,km\,s^{-1}$\\
&&&\\
\hline 
&&& \\ 
$\rho_{cloud} = $&$430\,cm^{-3}$&$1130\,cm^{-3}$&$31\,cm^{-3}$\\
 
&$23\,km\,s^{-1}$&$335\,km\,s^{-1}$&$6\,km\,s^{-1}$\\
 
$30\,cm^{-3}$&$ 8,900\,K$&$9,600\,K$&$23,500\,K$\\
&$\tau = 1.75$&$\tau = 0.34$&$\tau = 0.06$\\ 
&&&\\ 
\hline 
&&&\\
$\rho_{cloud} =$&& $1170\,cm^{-3}$ &$390\,cm^{-3}$\\ 
 
& &$100\,km\,s^{-1}$ &$380\,km\,s^{-1}$ \\

$60\,cm^{-3}$& &$8,300\,K$&$21,500\,K$\\ 
&$\tau = 4.65$&$\tau = 0.93$&$\tau = 0.17$\\ 
&&&\\
\hline 
&&$\hfill $&\\
$\rho_{cloud} =$&&$1360\,cm^{-3}$&$1690\,cm^{-3}$\\ 

& &$170\,km\,s^{-1}$&$722\,km\,s^{-1}$\\

$120\,cm^{-3}$& &$ 10,500\,K$&$ 11,300\,K$\\ 
&$\tau = 13.2$&$\tau = 2.64$&$\tau = 0.5$\\ 
&&&\\
\hline 
\end{tabular}\par}
\vspace{0.30cm}

%% file: tabella1.tex
{\centering \begin{tabular}{|c|c|c|c|}
\hline 
&\\
&$0.3 < \tau < 0.55 $\\
&\\
\hline
&\\ 
$\rho_{cloud} = 30\,cm^{-3}$&$4,000\,km\,s^{-1} < v_{jet} <7,500\,km\,s^{-1} $\\ 
&\\ 
\hline
&\\ 
$\rho_{cloud} = 60\,cm^{-3}$&$11,000\,km\,s^{-1} < v_{jet} <20,500\,km\,s^{-1} $\\ 
&\\ 
\hline
&\\ 
$\rho_{cloud} = 120\,cm^{-3}$&$30,000\,km\,s^{-1} < v_{jet} <55,500\,km\,s^{-1} $\\ 
&\\ 
\hline
\end{tabular}\par}
\vspace{0.30cm}

%% file: tabella2.tex
\vspace{0.30cm}
\hspace{17.cm}
{\centering \begin{tabular}{|c|c|c|c|}
\hline 
&&&\\
 &$v_{jet} = 1,300\,km\,s^{-1}$&$v_{jet} = 6,500\,km\,s^{-1}$&$v_{jet} = 
32,500\,km\,s^{-1}$\\
 &&&\\
\hline
&&&\\
$P_{kin}$ & $1.1\,10^{40}\,erg\,s^{-1}$ & $1.38\,10^{42}\,erg\,s^{-1}$ &
$1.73\,10^{44}\,erg\,s^{-1}$ \\
&&&\\
\hline 
&&& \\ 
&${(P_{rad})}_{max} = 0.03\,10^{40}\,erg\,s^{-1}$ & ${(P_{rad})}_{max} = 
0.02\,10^{42}\,erg\,s^{-1}$&${(P_{rad})}_{max} = 0.07\,10^{42}\,erg\,s^{-1}$ \\
 $\rho_{cloud} = 30\,cm^{-3}$& & &\\ 
  &$\eta_{max} = 2.7\%,\;\eta_{2t_{cc}} = 0.08\% $&$\eta_{max} = 
1.6\%,\;\eta_{2t_{cc}} = 0.4\% $
  &$\eta_{max} = 0.04\%,\;\eta_{2t_{cc}} = 0.5\,10^{-2}\% $\\ 
&&&\\ 
\hline 
&&&\\
 &  & ${(P_{rad})}_{max} = 0.14\,10^{42}\,erg\,s^{-1}$&${(P_{rad})}_{max} = 
0.15\,10^{42}\,erg\,s^{-1}$ \\
 $\rho_{cloud} = 60\,cm^{-3}$& & &\\ 
  & &$\eta_{max} = 10.\%,\;\eta_{2t_{cc}} = 0.5\% $&$\eta_{max} = 
0.09\%\,;\eta_{2t_{cc}} = 4.\,10^{-2}\% $\\ 
&&&\\
\hline
 &&&\\
 &  & ${(P_{rad})}_{max} = 0.02\,10^{42}\,erg\,s^{-1}$&${(P_{rad})}_{max} = 
0.35\,10^{42}\,erg\,s^{-1}$ \\
 $\rho_{cloud} = 120\,cm^{-3}$& & &\\ 
  & &$\eta_{max} = 1.4\%,\;\eta_{2t_{cc}} = 0.5\% $&$\eta_{max} = 
0.2\%,\;\eta_{2t_{cc}} = 1.\,10^{-2}\% $\\ 
&&&\\
\hline
\end{tabular}\par}
\vspace{0.30cm}